\documentclass[preprint,aip,jcp,a4,10pt,twocolumn,amsmath,amssymb]{revtex4-1}

\pdfoutput=1

\usepackage{bm}
\usepackage{graphicx}
\usepackage{natbib}

\usepackage{color}

\begin{document}

\title{Dynamics of a deformable active particle under shear flow}%

\author{Mitsusuke Tarama}
\affiliation{Department of Physics, Kyoto University, Kyoto 606-8502, Japan}
\affiliation{Institut f\"ur Theoretische Physik II: Weiche Materie, Heinrich-Heine-Universit\"at D\"usseldorf, D-40225 D\"usseldorf, Germany}
\affiliation{Institute for Solid State Physics, The University of Tokyo, Kashiwa, Chiba 277-8581, Japan}
\affiliation{Department of Physics, Graduate School of Science, The University of Tokyo, Tokyo 113-0033, Japan}

\author{Andreas M. Menzel}
\affiliation{Institut f\"ur Theoretische Physik II: Weiche Materie, Heinrich-Heine-Universit\"at D\"usseldorf, D-40225 D\"usseldorf, Germany}
\affiliation{Department of Physics, Kyoto University, Kyoto 606-8502, Japan}

\author{ Borge ten Hagen}
\affiliation{Institut f\"ur Theoretische Physik II: Weiche Materie, Heinrich-Heine-Universit\"at D\"usseldorf, D-40225 D\"usseldorf, Germany}

\author{Raphael Wittkowski}
\affiliation{SUPA, School of Physics and Astronomy, University of Edinburgh, Edinburgh, EH9 3JZ, United Kingdom}

\author{Takao Ohta}
\affiliation{Department of Physics, Kyoto University, Kyoto 606-8502, Japan}
\affiliation{Department of Physics, Graduate School of Science, The University of Tokyo, Tokyo 113-0033, Japan}
\affiliation{Soft Matter Center, Ochanomizu University, Tokyo 112-0012, Japan}

\author{Hartmut L\"owen}
\affiliation{Institut f\"ur Theoretische Physik II: Weiche Materie, Heinrich-Heine-Universit\"at D\"usseldorf, D-40225 D\"usseldorf, Germany}

\date{\today}

\begin{abstract}
The motion of a deformable active particle in linear shear flow is explored theoretically. 
Based on symmetry considerations, in two spatial dimensions, we propose coupled nonlinear
dynamical equations for the particle position, velocity, deformation, and rotation.
 In our model, both, passive rotations induced by the shear flow
as well as active spinning motions, are
taken into account. Our equations reduce to known models in the two limits of vanishing shear flow
and vanishing particle deformability. For varied shear rate and particle propulsion speed,
we solve the equations numerically and obtain a manifold of different dynamical modes
including active straight motion, periodic motions, motions on undulated cycloids, winding motions, as well as quasi-periodic and chaotic motions
induced at high shear rates. The types of motion are  distinguished
by different characteristics in the real-space trajectories and in the dynamical behavior 
of the particle orientation and its deformation.  Our predictions can be verified in experiments 
on self-propelled droplets exposed to a linear
shear flow.
\end{abstract}

\maketitle

\section{Introduction} \label{sec:Introduction}

In the last decade, the motion  and modeling of active particles has 
attracted much attention in the field of nonequilibrium physics 
\cite{Ebbens2010,Ramaswamy2010,Cates2012,Marchetti,Schimansky-Geier2012}. 
A major part of active particles are 
artificial colloidal microswimmers with fixed  stable shapes 
\cite{Paxton,Jiang2010,Kapral2013}, 
but there are also cases
in which the particles are deformable 
and do change shape during their motion. Such deformability is of basic importance
for active droplets 
\cite{Boukellal2004,Nagai2005,Toyota2009} 
but is also relevant for living swimmers like 
protozoa 
and other microorganisms 
\cite{Keren2009,Bosgraaf2009,Li2008,Maeda2008}.
Therefore a basic theoretical description for active deformable self-propelled particles and microswimmers is needed.

In a quiescent solvent, dynamical equations of motion were recently  put forward 
which couple the particle position and deformability 
\cite{OhtaOhkuma2009,Hiraiwa2010,Hiraiwa2011,Tarama2012,Tarama2013,Tarama2013PRE,Wada2009,Nishimura2009,Shao2010,Kruse2011,Ziebert2012}. 
One of the unexpected results was a spontaneous circling motion due to the coupling of deformability and self-propulsion
\cite{OhtaOhkuma2009,Hiraiwa2010,Hiraiwa2011}. 
However, in most practical situations 
\cite{Clement,Peyla,Bagorda2008},
various external fields are present to influence the particle motion. 
They are, for instance, induced by a chemoattractant, 
phototaxis, and gravity \cite{Friedrich2007,Jekely2009,Cates2009,Tarama2011,Wittkowski2012,Hiraiwa2013},
or external walls \cite{DiLuzio2005,Teeffelen2008,Wensink2008,Drescher2011}. 
An important particular case is a 
solvent flow field such as a Couette flow with a constant shear gradient or a Poiseuille flow
through tubes. There are several studies of rigid self-propelled particles in various shear geometries
\cite{Kessler1985,Kapral2010,Hagen2011,Stark2012}. However,
despite its practical relevance,  the motion of a deformable self-propelled particle in a solvent flow 
has not been considered theoretically yet. 
The corresponding modeling is expected to be complex
since  already rigid (undeformable) active particles have been shown to perform periodic
motion on cycloids (rather  than on straight lines) once they are exposed to a linear shear flow field
\cite{Hagen2011}.

In this paper we close this gap and propose a theoretical model for the motion of
an active deformable particle
in shear flow. We use symmetry considerations  in two spatial dimensions to obtain coupled nonlinear
dynamical equations for the particle position, velocity, deformation, and its active rotations.
 In our model, a passive rotation induced by the shear flow
and an active spinning motion are both
taken into account. On the one hand, for vanishing shear flow, our equations reduce to 
previous models for deformable particles \cite{OhtaOhkuma2009,Tarama2012,Tarama2013}. On the other hand, for vanishing particle deformability, we obtain the cycloidal motion
as embodied in previous investigations \cite{Hagen2011}. For varied shear rate and particle propulsion speed,
we solve the equations numerically and obtain a manifold of different dynamical modes
including active straight motion, periodic motions, regular and undulated cycloidal motions, winding motions, as well as quasi-periodic and chaotic motions 
induced at high shear rates.  The types of motion are  distinguished
by different characteristics in the dynamical behavior 
of the particle positions, velocity orientations, and its deformations.   We are not aware of any experiments of deformable active particles in shear flow, but in principle these experiments are conceivable building upon recent analysis of self-propelled
droplets  
\cite{Nagai2005,Takabatake2011}
 that can be subjected to an additional 
shear flow.

The organization of this paper is as follows. 
In the next section, we introduce time-evolution equations for an active deformable particle under an external flow. 
In Sec.~\ref{sec:Solution without deformation in 2d}, 
we consider the special case of a round disk-shaped non-deformable active particle by eliminating the variable for deformation. We relate this case to previous work \cite{Hagen2011}.
Next, in Sec.~\ref{sec:zeta=-0.1}, 
we present numerical 
results for the dynamics of a deformable active particle under steady shear flow. In Sec.~\ref{sec:zeta=1.5}, 
a spontaneous particle rotation for the dynamics is additionally included and numerical results are presented.
Finally, Sec.~\ref{sec:discussion} 
is devoted to a summary and to conclusions.

\section{Coupled Dynamical Equations} \label{sec:Model}

Based on symmetry considerations, we now derive a  set of coupled nonlinear dynamical equations to describe the motion of
 a deformable active particle under an externally imposed flow field. These equations are first listed for the general case of three spatial dimensions and an unspecified flow field. Afterwards we will confine ourselves to a two-dimensional geometry and consider a simple shear flow. As a first approach, we only investigate the influence of the shear flow on the particle dynamics and do not consider the inverse effect. 

In the following, we denote the prescribed externally imposed flow field that the particle is exposed to as $\bm{u}$. It is a given function of space. To proceed as normal 
\cite{Pleiner2002,Lubensky2003},
the elongational part of the fluid flow is extracted by the symmetric second-rank tensor $\mathbf{A}$ with components 
\begin{equation}
A_{ij} = \frac{1}{2} \left( \partial_i u_j +\partial_j u_i \right) 
 \;.
\label{eq:A}
\end{equation}
Similarly, the rotational part is extracted via the anti-symmetric second-rank tensor $\mathbf{W}$ with components
\begin{equation}
W_{ij} = \frac{1}{2} \left( \partial_i u_j -\partial_j u_i \right) 
 \;.
\label{eq:W}
\end{equation}

We denote the center-of-mass position of the particle at time $t$ as $\mathbf{x}(t)$. 
In general, the total particle velocity ${d {\bm {x}}}/{dt}$ has two contributions. On the one hand, the particle is ``passively'' advected by the externally imposed prescribed flow field $\bm{u}$. On the other hand, the particle can ``actively'' self-propel with respect to the surrounding fluid. The corresponding ``active'' velocity measured relatively to the surrounding fluid flow is denoted as $\mathbf{v}$. Altogether, we obtain the equation of motion
\begin{equation}
\frac{d x_i}{dt}= u_i +v_i
 \;,
\label{eq:x}
\end{equation}
where the index $i=1,2,3$ labels the Cartesian coordinates. 

In our description, the active velocity $\mathbf{v}$ is one of our major dynamical variables describing the behavior of the particle. The other ones are its deformation that we characterize by the second-rank traceless symmetric tensor $\mathbf{S}$, and an ``active'' particle rotation described 
by the second-rank antisymmetric tensor $\mathbf{\Omega}$. Both of these tensors are briefly introduced in the following. 

For simplicity, we only include elongational and flattening deformations of the particle. The tensor $\mathbf{S}$ represents these deformations 
\cite{OOS2009,Shitara2011}. 
We first consider the two-dimensional case, where orientations in the two-dimensional plane can be parameterized by a single angle $\varpi$. 

This angle $\varpi$ is now used to measure directions from the particle center of mass. 
The distance from the particle center to its boundary in the direction $\varpi$ at time $t$ is denoted as $R(\varpi,t)$. Large deformations are not taken into account, so that $R(\varpi,t)$ is single-valued with respect to the angle $\varpi$. 

For a steady circular shape we have $R(\varpi,t)=R_0$, with $R_0$ the particle radius. We now consider deviations $\delta R(\varpi,t)$ from the circular shape, such that the distance from the particle center to its boundary becomes $R(\varpi,t)=R_0+\delta R(\varpi,t)$. 
Next, the deviation from the circular shape $\delta R(\varpi,t)$ is expanded into a Fourier series 
\begin{equation}
\delta R(\varpi,t) = \sum_{m=2}^{\infty} \left( z_{m}(t) e^{i m \varpi} +z_{-m}(t) e^{-i m \varpi} \right)
 \;.
 \label{eq:0.6}
\end{equation}
In this expansion, the zeroth mode is excluded by assuming that the area of the particle is conserved. The first Fourier mode would represent a translation of the center of mass, which we already took into account by the velocity variable $\mathbf{v}$. 
Therefore, the lowest mode describing deformations is the second one, which actually represents an elliptical deformation. 
In two dimensions, we can define a symmetric tensor as $S_{11}=-S_{22}=z_2+z_{-2}=s \cos 2 \theta$ and $S_{12}=S_{21}=i(z_2-z_{-2})=s \sin 2\theta$, where we have defined $z_{\pm2}=(s/2) e^{\mp 2 i \theta}$. 
So the searched-for tensor $\mathbf{S}$ can be written in the form
\begin{equation}\label{S}
\mathbf{S}=\left(\begin{array}{cc}
S_{11} & S_{12} \\
S_{12} & -S_{11}
\end{array}\right)
=s\left(\begin{array}{cc}
\cos 2\theta & \sin 2\theta \\
\sin 2\theta & -\cos 2\theta
\end{array}\right).
\end{equation}
Here, $s$ corresponds to the degree of deformation, and $\theta$ to the orientation of the long axis of deformation. 
\begin{eqnarray}
R(\varpi,t) 
 &=& R_0 +s(t) \cos 2 \left[ \varpi -\theta(t) \right] \notag\\
 &=& R_0 +S_{11}(t) \cos 2 \varpi +S_{12}(t) \sin 2 \varpi
 \;.
 \label{eq:0.7}
\end{eqnarray}

In the case of three spatial dimensions, the deviation $\delta R$ must be expanded into spherical harmonics $ Y_{\ell m} (\tilde{\varpi})$ with coefficients $c_{\ell m} (t)$ and  $\tilde{\varpi}$ the solid angle.  
Likewise, the minimum mode of the deformation, $\ell=2$, represents an ellipsoidal deformation. 
See Ref.~\citenum{Hiraiwa2011} 
for the relations between $c_{\ell m}$ and $\mathbf{S}$ in three spatial dimensions.

Finally, our last dynamic variable $\mathbf{\Omega}$ characterizes an ``active'' rotational motion of the particle around its center of mass. We call this a spinning motion 
\cite{Tarama2012,Tarama2013}. 
This antisymmetric second-rank tensor $\mathbf{\Omega}$ can be obtained from the corresponding vector of angular velocity $\bm{\omega}$ via 
\cite{Tarama2013}
\begin{equation}
\Omega_{ij} = \epsilon_{ijk} \omega_k 
 \;,
 \label{eq:omega}
\end{equation}
where $\epsilon_{ijk}$ denotes the components of the Levi-Civita tensor. 
Summation over repeated indices is implied, as throughout the remaining part of this paper. 

The spinning motion $\mathbf{\Omega}$ occurs in addition to the rotational motion prescribed by the external flow field $\mathbf{u}$, see Eq.~(\ref{eq:W}). In other words, the anti-symmetric tensor $\mathbf{\Omega}$ represents the relative rotation with respect to the rotational motion $\mathbf{W}$ of the surrounding fluid flow. 
Therefore, $\mathbf{W}+\mathbf{\Omega}$ describes the total angular velocity with respect to the laboratory frame from which the flow field is parameterized. 

In total, we have introduced three central dynamical variables to characterize the state of a deformable active particle: $\mathbf{v}$ for the active propulsion velocity, $\mathbf{S}$ for the particle deformation, and $\mathbf{\Omega}$ for the active rotational motion. 
Based on symmetry arguments, our model for the dynamic evolution of these variables is derived. We consider the following set of coupled nonlinear equations:
\begin{eqnarray}
\lefteqn{\frac{d v_i}{dt} +a_2 \left( W_{ik} + \Omega_{ik} \right) v_k =} \notag\\
&& \qquad\alpha v_i -(v_k v_k)v_i -a_1 S_{ik}v_k 
 \;, 
\label{eq:v}
\\[.1cm]
\lefteqn{\frac{d S_{ij}}{dt} - b_2 \left[ S_{ik} \left( W_{kj} +\Omega_{kj} \right) - \left( W_{ik} +\Omega_{ik} \right) S_{kj} \right] =}  \notag \\
 &&{} \qquad-\kappa S_{ij}+b_1 \left[ v_i v_j -\frac{\delta_{ij}}{d} \left( v_k v_k \right) \right] 
 + b_3 \Omega_{ik} S_{k\ell} \Omega_{\ell j}   \notag\\ 
 &&{} \qquad+ b_4 \Omega_{k\ell} \Omega_{k\ell} S_{ij}   
 + \nu_1 \left[ A_{ij}-\frac{\delta_{ij}}{d} A_{kk} \right] \notag\\
 &&{} \qquad+ \nu_2 \left[ A_{ik} S_{kj} +S_{ik} A_{kj} -\frac{2 \delta_{ij}}{d} S_{k \ell} A_{\ell k} \right]
 \;,
 \label{eq:S}
\\[.1cm]
\lefteqn{\frac{d \Omega_{ij} }{d t} = \zeta\Omega_{ij} +\Omega_{ik} \Omega_{k\ell} \Omega_{\ell j} }
\notag\\
&&{} \qquad+c_1 \left( S_{ik} \Omega_{kj} +\Omega_{ik} S_{kj} \right) 
  + c_2 S_{ik} \Omega_{k\ell} S_{\ell j}
 \;.
 \label{eq:Omega}
\end{eqnarray}
Here $\delta_{ij}$ denotes the Kronecker delta, and $d$ is the dimension of space. The coefficients $\alpha$, $\kappa$, 
$\zeta$,
$a_1$, $a_2$, $b_1$, $b_2$, $b_3$, $b_4$, $c_1$, $c_2$, $\nu_1$, and $\nu_2$ are phenomenological coupling parameters. 
We now comment on each of the terms in this set of equations for the time evolution, Eqs.~(\ref{eq:v})--(\ref{eq:Omega}). In principle, more terms and higher-order couplings can be included, but the current model covers the main physical aspects that we intend to describe. 

We start with the first two terms on the right-hand side of Eq.~(\ref{eq:v}). They can be rewritten as
\begin{equation}
- \frac{\partial F}{\partial v_i} \quad\mbox{with}\quad F = - \frac{\alpha}{2} \left( v_k v_k \right) +\frac{1}{4} \left( v_k v_k \right)^2
 \;,
 \label{eq:0.2}
\end{equation}
where $F$ is a Lyapunov function controlling the spontaneous self-propulsion. With increasing $\alpha$, $F$ describes a bifurcation at $\alpha=0$ corresponding to the onset of active motion with $\mathbf{v}\neq\mathbf{0}$. 
In the same way, the first two terms on the right-hand side of Eq.~(\ref{eq:Omega}) can be rewritten as  
\begin{equation}
-\frac{\partial G}{\partial \Omega_{ij}} \quad\mbox{with}\quad G = \frac{\zeta}{2} {\rm tr}\ \Omega^2 +\frac{1}{4} {\rm tr}\ \Omega^4
 \;,
 \label{eq:0.3}
\end{equation}
introducing another Lyapunov function $G$. 
Likewise, this function characterizes the onset of the spontaneous rotation of the particle around its center of mass when $\zeta$ becomes positive. 
Together, the coefficients $\alpha$ and $\zeta$ characterize the strength of activity, for self-propulsion and for active rotation, respectively. 
In contrast to that, an active deformation of the particle is not considered. The first term on the right-hand side of Eq.~(\ref{eq:S}) with $\kappa>0$ always induces a relaxation of the deformation back to a spherical (circular) shape, at least when the coupling to $\mathbf{v}$, $\mathbf{\Omega}$, and to the surrounding flow field $\bm{u}$ allow it. 

Next, we consider the terms with the coefficients $a_2$ in Eq.~(\ref{eq:v}) and $b_2$ in Eq.~(\ref{eq:S}). They have similar origin and include reorientations of the particle velocity and elongation axes due to the shear flow and due to the active spinning motion. In the passive case, they would contain the advective reorientation of the particle axes due to the fluid flow. Since an active particle can follow a prescribed rule on how to react to external rotational flow fields, the numerical values of the coefficients cannot be generally fixed at this point. We assume $a_2>0$. In principle, this contribution with $a_2>0$ can describe a sort of Magnus effect, with a force acting onto the particle in the direction perpendicular to its velocity and angular velocity. 
For a rigid particle -- i.e.\ an undeformable particle -- of spherical or ellipsoidal shape, rotational and translational motions do not couple to each other to linear order \cite{Doi2004}. Accordingly, the coupling between the velocity $\mathbf{v}$ and the rotational part of the flow field $\mathbf{W}$ is nonlinear in the $a_2$-term.

The third term on the right-hand side of Eq.~(\ref{eq:v}) with the coefficient $a_1$ and the second term on the right-hand side of Eq.~(\ref{eq:S}) with the coefficient $b_1$ are the leading-order coupling terms between 
the velocity $\mathbf{v}$ and the deformation $\mathbf{S}$. Their influence was already extensively studied in previous investigations of deformable self-propelled particles 
\cite{OhtaOhkuma2009,Hiraiwa2011}. 
On the one hand, deformations can reorient the particle velocity and change its speed via the $a_1$-term. Bended particle trajectories can result from this contribution. 
On the other hand, via the $b_1$-term, deformations can be 
induced when the particle self-propels.

In addition to that, we include further coupling contributions between the deformation $\mathbf{S}$ and the active rotation $\mathbf{\Omega}$ \cite{Tarama2012,Tarama2013}. These are the terms with the coefficients $b_3$ and $b_4$ on the right-hand side of Eq.~(\ref{eq:S}), and the terms with the coefficients $c_1$ and $c_2$ on the right-hand side of Eq.~(\ref{eq:Omega}). 
For $b_3>0$ and $b_4>0$, the self-driven active rotation 
in a two-dimensional space enhances 
the degree of deformation, while it 
reduces 
it for $b_3<0$ and $b_4<0$.
When we confine ourselves to two spatial dimensions in the following sections, these two terms are equivalent for $b_3=2 b_4$. 
We remark that, in three spatial dimensions, the term with the coefficient $b_3$ has an additional effect to rotate the particle, in contrast to the $b_4$-term. 
The third and fourth terms 
on the right-hand side of Eq.~(\ref{eq:Omega}) include the analogous effects on the rotations $\mathbf{\Omega}$ induced by the deformation $\mathbf{S}$.  
Here, note that the $c_1$ term vanishes in two dimensions. 
Again, since different sorts of active particles may feature different coupling properties between their deformations and rotations, the values of these coefficients depend on the system under consideration. 

Finally, the elongational part of the externally imposed flow field can lead to a deformation of the particle. This effect is included by the last two contributions on the right-hand side of Eq.~(\ref{eq:S}). The corresponding coefficients $\nu_1$ and $\nu_2$ describe how the active particle reacts to the straining part of the flow field. 
Our terms are consistent 
with those of a previous study on the dynamics of a non-active liquid droplet in a fluid flow, where also elliptical shape deformations were considered\cite{Maffettone1998,Lubensky2003}. 
We note that the contribution with the coefficient $\nu_2$ vanishes for a two-dimensional geometry of incompressible flow. Such a case is studied below. 

In contrast to Eq.~(\ref{eq:S}), the tensor $\mathbf{A}$ containing the elongational part of the fluid flow does not enter Eq.~(\ref{eq:v}) for the velocity $\mathbf{v}$. This is because $\mathbf{v}$ is defined as the {\it relative} 
velocity with respect to the fluid flow $\mathbf{u}$, see Eq.~(\ref{eq:x}). Likewise, the rotational part of the fluid flow characterized by the tensor $\mathbf{W}$ is absent in Eq.~(\ref{eq:Omega}): $\mathbf{\Omega}$ describes the {\it relative} 
rotation with respect to the 
surrounding flow field. 
In principle, coupling terms between $\mathbf{\Omega}$ and the tensor $\mathbf{A}$ are possible. This would mean that the active particle features a way of reacting to an elongational flow by adjusting its spinning motion. 
However, we do not consider such a process. 

Generally our equations apply to a three-dimensional set-up. For simplicity, however, we confine ourselves to two spatial dimensions for the remaining part of this paper. Furthermore, we from now on specify the externally imposed flow field $\mathbf{u}$ to a linear steady shear flow, 
\begin{equation}
{\bf u}=(\dot{\gamma} y, 0 ) \;,
\label{eq:ugamma}
\end{equation} 
with $\dot{\gamma}$ the shear rate.

\section{Dynamics without deformation} \label{sec:Solution without deformation in 2d}

The full set of dynamic equations (\ref{eq:x}), (\ref{eq:v})--(\ref{eq:Omega}) is very complex. To get a first overview, we start by studying a reduced model. More precisely, we neglect deformability, i.e.\ we set $\mathbf{S}=\mathbf{0}$, and consider a circularly-shaped rigid particle. Under certain assumptions, an analytical solution can be obtained in this case. 

Prescribing $\mathbf{S}=\mathbf{0}$, Eq.~(\ref{eq:S}) is dropped from our system of equations. Eqs.~(\ref{eq:x}), (\ref{eq:v}), and (\ref{eq:Omega}) reduce to
\begin{eqnarray}
\frac{d x_i}{d t} 
 &=& v_i +u_i
 \;,
 \label{eq:1.0}
\\
\frac{d v_i}{d t} 
 &=&  \alpha v_i -\left( v_k v_k \right) v_i -a_2 \left( W_{ik} + \Omega_{ik} \right) v_k
 \;,
 \label{eq:1.1}
\\
\frac{d \Omega_{ij}}{d t}  &=& 
\zeta\Omega_{ij}+\Omega_{ik}\Omega_{kl}\Omega_{lj}
 \;.
 \label{eq:1.2}
\end{eqnarray}
We parameterize the vectors and tensors by $\mathbf{x} = (x, y)$, 
\begin{equation}\label{v}
\mathbf{v} = (v \cos \phi, v \sin \phi ), 
\end{equation}
$\Omega_{11}=\Omega_{22}=0$, as well as $\Omega_{12} = -\Omega_{21} = \omega$, and insert $\mathbf{W}$ using Eqs.~(\ref{eq:W}) and (\ref{eq:ugamma}). 
Then Eqs.~(\ref{eq:1.0})--(\ref{eq:1.2}) become
\begin{eqnarray}
\frac{d x}{d t} &=& v \cos \phi +\dot{\gamma} y
 \;,
 \label{eq:1.8}\\
\frac{d y}{d t} &=& v \sin \phi 
 \;,
 \label{eq:1.9}\\
\frac{d v}{d t} 
 &=& \alpha v -v^3 
 \;,
 \label{eq:1.5}\\
\frac{d \phi}{d t} 
 &=& a_2 \left( - \frac{\dot{\gamma}}{2} +\omega  \right) 
 \;,
 \label{eq:1.6}\\ 
\frac{d \omega}{d t} 
 &=&  \zeta \omega -\omega^3
 \;.
 \label{eq:1.7}
\end{eqnarray}

Next, we assume that the magnitudes of the velocity $v$ and of the relative rotation $\omega$ relax quickly, so that they are given by the steady state solutions of Eqs.~(\ref{eq:1.5}) and (\ref{eq:1.7}), respectively. 
In this situation, together with $\alpha>0$ and $\zeta>0$ implying self-propulsion and active spinning, we have 
\begin{equation}
v = \sqrt{\alpha} 
 \;,
 \label{eq:1.10}
\end{equation}
\begin{equation}
\omega = \pm \sqrt{\zeta} 
 \;,
 \label{eq:1.11}
\end{equation}
where the positive and negative signs in Eq.~(\ref{eq:1.11}) correspond to counter-clockwise and clockwise rotations, respectively. 
Using these solutions, Eq.~(\ref{eq:1.6}) reads
\begin{equation}
\frac{d \phi}{d t}  = a_2 \left( -\frac{\dot{\gamma}}{2} \pm\sqrt{\zeta} \right)
 \;.
 \label{eq:1.12}
\end{equation}
From Eqs.~(\ref{eq:1.8}), (\ref{eq:1.9}), (\ref{eq:1.10}), and (\ref{eq:1.12}), the trajectory of the center of mass can be calculated as
\begin{eqnarray}
x(t) 
 &=& \frac{ \sqrt{\alpha} \left\{ a_2 \left( -\frac{\dot{\gamma}}{2} \pm\sqrt{\zeta} \right) -\dot{\gamma} \right\} }{a_2^2 \left( -\frac{\dot{\gamma}}{2} \pm\sqrt{\zeta} \right)^2} \left\{ \sin [\phi(t)] -\sin \phi_0 \right\} \notag\\
 &&{}+ \dot{\gamma} \left( \frac{ \sqrt{\alpha}}{a_2 \left( -\frac{\dot{\gamma}}{2} \pm\sqrt{\zeta} \right)} \cos \phi_0 +y_0 \right) t  +x_0
 \;,
 \label{eq:1.13}\\
y(t) &=& \frac{ - \sqrt{\alpha} }{a_2 \left( -\frac{\dot{\gamma}}{2} \pm\sqrt{\zeta} \right)} \left\{ \cos [\phi(t)] -\cos \phi_0 \right\} + y_0
 \;,
 \label{eq:1.14}\\
\phi(t) &=& a_2 \left( -\frac{\dot{\gamma}}{2} \pm\sqrt{\zeta} \right) t +\phi_0 
 \;.
 \label{eq:1.17}
\end{eqnarray}
Here, $( x_0, y_0)$ and $\phi_0$ are the position of the center of mass and the direction of the velocity vector at $t=0$, respectively. 
This set of solutions represents a cycloidal trajectory. 

A similar cycloidal trajectory has previously been obtained for an active rigid circularly-shaped particle by some of the present authors \cite{Hagen2011}. 
In that case, the dynamics of the particle features a polarity axis 
\cite{Hagen2011JPCM}, 
the orientation of which in the two-dimensional plane can be characterized by the angle $\phi$.  
It marks the direction of the self-propulsion that generates a relative velocity with respect to the surrounding flow field. 
The equations of motion introduced in Ref.~\citenum{Hagen2011} 
can be written in the form
\begin{eqnarray}
\frac{d x}{d t} &=& \dot{\gamma} y + \tilde{\alpha} \left[\cos \phi + f_x \right]
 \;,
 \label{eq:Hagen.1}\\
\frac{d y}{d t} &=& \tilde{\alpha} \left[\sin \phi + f_y \right]
 \;,
 \label{eq:Hagen.2}\\
\frac{d \phi}{d t} &=& -\frac{\dot{\gamma}}{2} +\tilde{\mu}(1 + g)
 \;,
 \label{eq:Hagen.3}
\end{eqnarray}
where $\tilde{\alpha}$ is a normalized effective self-propulsion force that is proportional to the self-propulsion 
velocity $v$ in the over-damped regime considered in Ref.~\citenum{Hagen2011}. 
$\tilde{\mu}$ accounts for an additional self-induced 
\cite{Kummel2013} 
or externally imposed 
\cite{Baraban2012} 
torque on the particle.
Equations (\ref{eq:Hagen.1})--(\ref{eq:Hagen.3}) 
contain Gaussian white noise terms $f_x$, $f_y$, and $g$, which are not included in the present approach. 
Our equations (\ref{eq:1.8}), (\ref{eq:1.9}), and (\ref{eq:1.12}), together with the asymptotic steady-state magnitudes 
of the translational and angular velocities in Eqs.~(\ref{eq:1.10}) and (\ref{eq:1.11}), respectively, are consistent with Eqs.~(\ref{eq:Hagen.1})--(\ref{eq:Hagen.3}) when the noise terms are neglected. By solving this zero temperature limit for $\tilde{\alpha}=\alpha^{1/2}$ and  $\tilde{\mu}=\pm \zeta^{1/2}$, one recovers the results presented in Eqs.~(\ref{eq:1.13})--(\ref{eq:1.17}) for $a_2=1$.

For the special case of $(\dot{\gamma}/2) =\pm \sqrt{\zeta} $, the solutions of Eqs.~(\ref{eq:1.8}), (\ref{eq:1.9}), and (\ref{eq:1.12}), together with Eq.~(\ref{eq:1.10}), read 
\begin{eqnarray}
x(t)
 &=& \left( \frac{\dot{\gamma}}{2} \sqrt{\alpha} \sin \phi_0 \right) t^2 \notag\\
 && {}+ \left( \sqrt{\alpha} \cos \phi_0 +\dot{\gamma} y_0 \right) t +x_0
 \;,
 \label{eq:1.15}\\
y(t)
 &=& \left( \sqrt{\alpha} \sin \phi_0 \right) t +y_0
 \;,
 \label{eq:1.16} \\
\phi(t) &=& \phi_0 
 \;.
\end{eqnarray}
The physical meaning of this limit is that the spontaneous rotation compensates the rotation due to the surrounding flow field, i.e. $(\dot{\gamma}/2 )= \pm \sqrt{\zeta} $ in Eq.~(\ref{eq:1.12}) or correspondingly $(\dot{\gamma}/2) =\tilde{\mu}$ in Eq.~(\ref{eq:Hagen.3}). 
Also Eqs.~(\ref{eq:1.15}) and (\ref{eq:1.16}) are consistent with the ones correspondingly obtained in Ref.~\citenum{Hagen2011}.

\section{Dynamics without active rotation} \label{sec:zeta=-0.1}

\begin{figure*}[tbhp]
  \begin{center}
         \includegraphics[width=7 in]{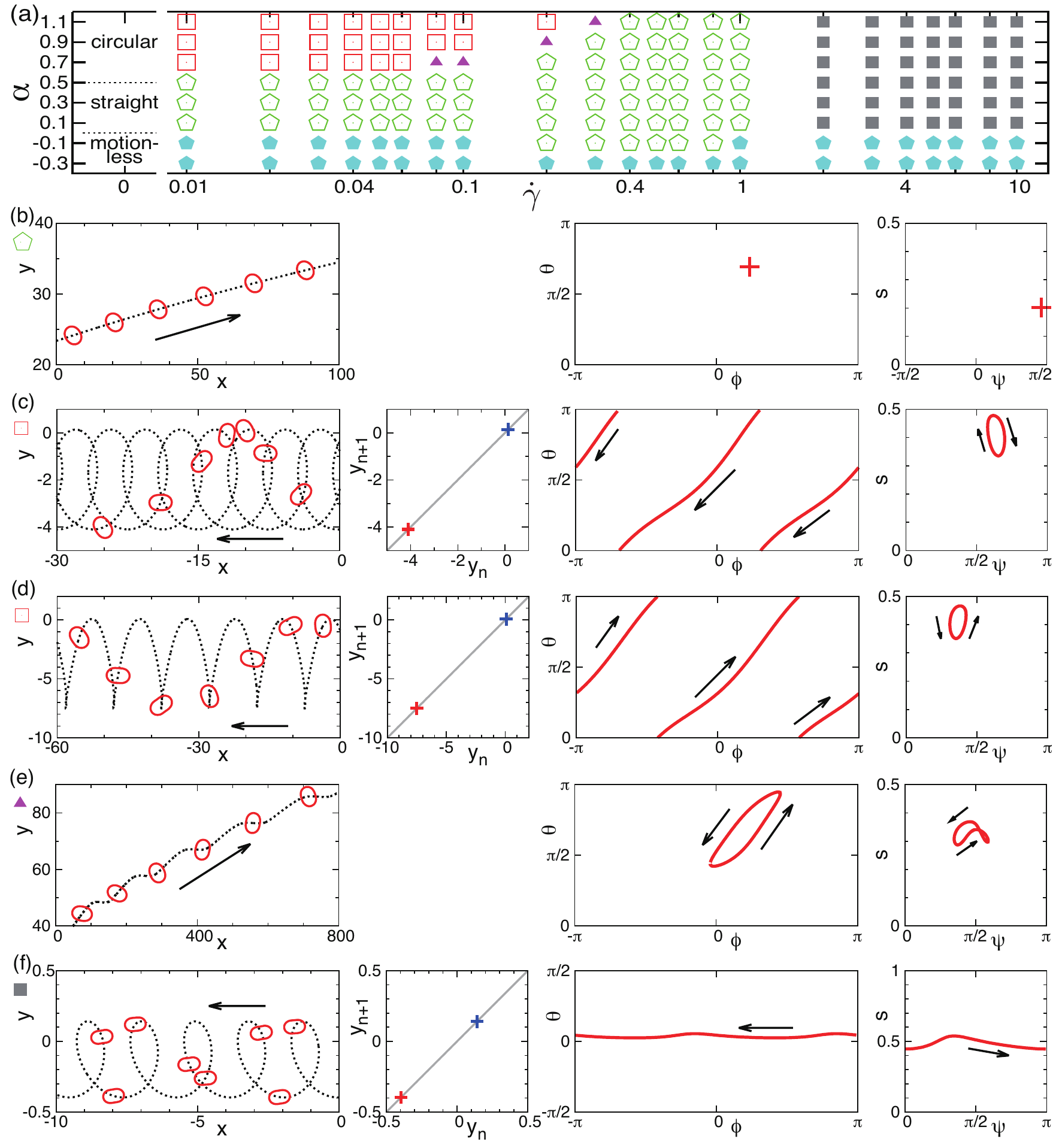}
      \caption{
      (a) Dynamical phase diagram and (b)--(f) trajectories in real space ($1$st column), return maps ($2$nd column), attractors in $\theta$-$\phi$ 
      space ($3$rd column) as well as in $s$-$\psi$ space ($4$th column) of the 
      typical dynamical motions, obtained by solving Eqs.~(\ref{eq:x}) and (\ref{eq:v})--(\ref{eq:Omega}) numerically in two dimensions without active rotational motion ($\zeta=-0.1$); 
      (b)       {\it active straight motion} 
      for $\alpha=0.5$ and $\dot{\gamma}=0.1$ indicated by the green open pentagons in panel~(a); 
      (c) and (d)       {\it cycloidal I motions} 
      of clockwise and counter-clockwise rotations of the particle deformations, respectively, for $\alpha=0.9$ and $\dot{\gamma}=0.1$ marked by the red open squares in panel~(a); 
      (e)       {\it winding I motion} 
      for $\alpha=0.7$ and $\dot{\gamma}=0.08$ indicated by the purple filled triangles in panel~(a); 
      (f)       {\it cycloidal II motion} 
      for $\alpha=0.1$ and $\dot{\gamma}=2$ marked by the gray filled squares in panel~(a). 
      Arrows in panels~(b)--(f) show the directions of motion. 
      Some snapshots of the particle, the size of which is adjusted for illustration, are superimposed to the trajectory
      in real space. 
      Turquoise 
      filled pentagons in panel~(a) 
      represent the       {\it passive straight motion} 
      with $v=0$. 
      The variable $\mathbf{\Omega}$ corresponding to active spinning equals $\mathbf{0}$ for all of these types of motion. 
      A return map for the {\it active straight motion} and {\it winding I motion} in panels~(b) and (e) does not exist because the $y$-component of the velocity does not change its sign in these motions. 
      } \label{fig:zeta-0.1}
  \end{center}
\end{figure*}

In the previous section, we studied a rigid non-deformable particle as a first step. We now include deformability, but do not consider an active spinning motion of the particle. Although various different dynamic states can be found in this situation, the dynamics is still much simpler than with active rotations included, as is shown later. The dynamic equations must be solved numerically. 

Choosing $\zeta<0$ hinders active spinning.  
To solve Eqs.~(\ref{eq:x}) and (\ref{eq:v})--(\ref{eq:Omega}), we use a fourth-order Runge-Kutta method.  
We checked the numerical accuracy by comparing results obtained for different time increments.

The full parameter space is far too complex to be exhaustively explored. We therefore concentrate on the impact of only two parameters that we consider central to the current problem. One of them is the strength of the self-propulsion of the particle characterized by the parameter $\alpha$. The other one is the strength of the imposed shear flow determined by the shear rate $\dot{\gamma}$. 

All coupling parameters are fixed at similar magnitude to allow an equal impact of the corresponding effects on the system behavior. We set $a_1=b_1=-1$, $a_2=b_2=c_2=\nu_1=1$, and $b_3 +2 b_4 = 1$ (as noted in Sec.~\ref{sec:Model} the terms with the coefficients $b_3$ and $b_4$ coincide in two spatial dimensions, and the terms with the coefficients $c_1$ and $\nu_2$ vanish in our geometry). 
Intermediate damping rates are used for the deformations and for the spinning motion by imposing $\kappa=0.5$ and $\zeta=-0.1$, respectively. 
We obtained our results by directly numerically integrating Eqs.~(\ref{eq:x}) and (\ref{eq:v})--(\ref{eq:Omega}). After that we reparameterized them for illustrative purposes using Eqs.~(\ref{S}) and (\ref{v}).

Our results are summarized in Fig.~\ref{fig:zeta-0.1}. 
We present in Fig.~\ref{fig:zeta-0.1}(a) a phase diagram in the parameter plane of the self-propulsion strength $\alpha$ and the shear rate $\dot{\gamma}$. Various qualitatively different types of dynamical states are found and explained in more detail below. They are indicated in the phase diagram by the different symbols. At the position included as $\dot{\gamma}=0$, we describe in words the type of motion observed at zero shear rate for the different self-propulsion strengths $\alpha$. Increasing the shear rate towards the right boundary of the phase diagram, we can see how the shear flow influences the dynamic behavior of the particle. 

Each of the observed dynamic states is characterized separately in the rows of Figs.~\ref{fig:zeta-0.1}(b)--(f). The location in the phase diagram is indicated by the corresponding symbol below the panel number.  We present typical real-space trajectories by the black dotted lines in the first column. Black arrows indicate the direction of migration. The trajectories are drawn from the laboratory frame. Therefore their appearance strongly depends on the initial $y$-coordinate: the advective flow velocity increases in $y$-direction due to the shear geometry and leads to a stretching of the trajectories in $x$-direction. 
In particular, the direction of motion also depends 
on the $y$-coordinate: the flow field points to the right for $y>0$, whereas it points to the left for $y<0$. To avoid confusion, we remark again that the dynamics that is described by Eqs.~(\ref{eq:v})--(\ref{eq:Omega}) itself is not affected in this way, because only the relative velocity $\mathbf{v}$ with respect to the shear flow is considered and only gradients of the flow velocity enter via the tensors $\mathbf{A}$ and $\mathbf{W}$. 

In order to illustrate the current state of deformation and orientation along the trajectories that are drawn in the first column of Figs.~\ref{fig:zeta-0.1}(b)--(f), representative snapshots of the particle are superimposed in red. 
For the purpose of best visualization, the size of the particle was adjusted and the deformations are not drawn as pure ellipsoids. To further characterize the modes of migration, return maps of the corresponding motion in real space are included in the second column of Figs.~\ref{fig:zeta-0.1}(b)--(f). We extract the local maximum and minimum values of $y$ along each real-space trajectory for several thousand up-down oscillations. The maximum and minimum values are labeled as $y_n$ ($n=1,2,\dots$) and plotted as blue and red points, respectively, in a return map $y_{n+1}$ vs.\ $y_n$. In other words, we calculated the return maps at the Poincar\'e sections where $v_y=0$ for $d v_y/dt < 0$ and $d v_y/dt > 0$, respectively. The diagonal line in the return maps is included for illustration and does not represent any data points. 

Finally, the attractors in phase space are drawn in the third and fourth columns. 
Black arrows indicate the direction of motion along the attractors. We show plots in $\theta$-$\phi$ space (third column) and in $s$-$\psi$ space (fourth column). 
As introduced 
in Eqs.~(\ref{S}) and (\ref{v}),
$\theta$ and $\phi$ describe the orientation of the long axis of the deformation tensor 
$\mathbf{S}$ and the orientation of the relative velocity vector $\mathbf{v}$, respectively. $\theta$ and $\phi$ are observed from the laboratory frame. This is different for $s$ and $\psi$. First, $s$ measures the magnitude of deformation as is obvious from Eq.~(\ref{eq:0.7}). Second, $\psi$ is defined as the relative angle between the long axis of deformation 
 and the velocity orientation, 
\begin{equation}\label{psi}
\psi=\theta-\phi. 
\end{equation}
Thus $s$ and $\psi$ are measured in the co-moving particle frame. 
Both attractors, in the $\theta$-$\phi$ space and in the $s$-$\psi$ space, are independent of the initial $y$-coordinate.

We now go through the different dynamic states depicted in Fig.~\ref{fig:zeta-0.1}.
The most trivial state is represented by the turquoise filled pentagon symbols in Fig.~\ref{fig:zeta-0.1}(a). They indicate a {\it passive straight motion}. 
In the absence of any external flow field, a particle in this state is motionless and has a circular shape. 
When the shear flow is switched on, the particle is elongated due to the elongational contribution from the flow field. Nevertheless, its active self-propulsion velocity remains zero, $v=0$, so that it is just passively advected with the flow field. We do not include plots of these trivial trajectories in Fig.~\ref{fig:zeta-0.1}.

Next, a particle that moves straight in a condition without shear flow continues to move straight in the presence of shear flow at low shear rates $\dot{\gamma}$. It features a time-independent steady state of deformation. Such a situation is marked by the green open pentagons in the phase diagram Fig.~\ref{fig:zeta-0.1}(a) and shown in Fig.~\ref{fig:zeta-0.1}(b) for $\alpha=0.5$ and $\dot{\gamma}=0.1$. The real-space trajectory is only bended a little because the particle is advected in $x$-direction with the fluid flow that increases in the $y$-direction due to the shear geometry. 
Since $v\neq0$, we term this type of motion an {\it active straight motion} in this paper. 
We note from the line $\alpha=-0.1$ in the phase diagram Fig.~\ref{fig:zeta-0.1}(a) that with increasing shear rate $\dot{\gamma}$ a transition from passive to active straight motion can be induced. Interestingly, the phase behavior is reentrant, and we again observe passive straight motion at very high shear rates. 
The reason for this behavior are shear-rate dependent deformations $\mathbf{S}$ that are induced in Eq.~(\ref{eq:S}) via the shear flow. They in turn couple to the relative velocity $\mathbf{v}$ in Eq.~(\ref{eq:v}) and at intermediate shear rates induce active self-propulsion.

If a particle undergoes a circular motion when the shear is absent, it exhibits what we call a {\it cycloidal I motion} 
under a small nonzero shear rate as indicated by the red open squares in Fig.~\ref{fig:zeta-0.1}(a). 
In this state, a particle moves on a cycloidal trajectory with $v\neq0$ and with its deformation axes rotating as depicted in Figs.~\ref{fig:zeta-0.1}(c) and (d), both for $\alpha=0.9$ and $\dot{\gamma}=0.1$. Both, clockwise (c) and counter-clockwise (d) rotations are possible. 
Whether clockwise or counter-clockwise rotation appears during the cycloidal I motion generally depends on the initial conditions. At higher shear rates close to the stability boundary of the cycloidal I motion, however, cycloidal I motion of 
counter-clockwise 
rotation becomes unstable first, before the one with 
clockwise rotation. 
This is because the rotational part of the shear flow is oriented in clockwise direction as well and breaks the rotational symmetry of space. 
However, the effect occurs within a thinner parameter region than the grid size in Fig.~\ref{fig:zeta-0.1}(a) resolves. Therefore we do not mark this region in the phase diagram Fig.~\ref{fig:zeta-0.1}(a).

So far, we have only discussed types of motion that result directly as a generalization of the types of motion found for vanishing flow field $\dot{\gamma}=0$ 
\cite{OhtaOhkuma2009,Tarama2013}. 
Quite contrarily, the following types of motion are qualitatively different and newly observed in the presence of the shear flow. 

When the cycloidal I motion has become unstable at high shear rates, the particle exhibits a {\it winding I motion}. 
The corresponding narrow region in the phase diagram Fig.~\ref{fig:zeta-0.1}(a) is marked by the purple filled triangles. It is located between the cycloidal I motion and the active straight motion. 
For this winding I motion, the long axis 
of the particle does not make full rotations in the laboratory frame. It only oscillates in time around the velocity vector, as shown in Fig.~\ref{fig:zeta-0.1}(e) for $\alpha=0.7$ and $\dot{\gamma}=0.08$. 
In particular, the trajectories in $\theta$-$\phi$ space 
exhibit a closed loop indicating an oscillation. 
In contrast to the active straight motion, both, the relative velocity and the deformations of the particle, are time-dependent. 

Finally, at high shear rates, also the active straight motion becomes unstable, and a {\it cycloidal II motion} 
appears. It is indicated by the gray filled squares in Fig.~\ref{fig:zeta-0.1}(a) and further characterized in Fig.~\ref{fig:zeta-0.1}(f) for $\alpha=0.1$ and $\dot{\gamma}=2$. 
Again the trajectory in real space is of cycloidal shape. 
However, as can be seen from the trajectory in $\theta$-$\phi$ space, the value of $\theta$ stays close to zero with only small oscillations around it. Thus, in contrast to the cycloidal I motion, the elongation axis of deformation remains approximately horizontal for all times.

\section{Full dynamics} \label{sec:zeta=1.5}

In the previous two sections, we considered simplified special cases of the dynamic equations (\ref{eq:x}) and (\ref{eq:v})--(\ref{eq:Omega}) to identify the basic states of motion. First we neglected deformations in Sec.~\ref{sec:Solution without deformation in 2d}, then we excluded active contributions from the rotational spinning motion in Sec.~\ref{sec:zeta=-0.1}. 
Nevertheless, the dynamics in both cases was already quite complex. This complexity is increased even further when we now investigate the full active dynamics. For example qualitatively new quasi-periodic and chaotic states arise. 

We use the same methods and the same parameter values as in the previous section to study the full set of dynamic equations (\ref{eq:x}) and (\ref{eq:v})--(\ref{eq:Omega}). The only difference is that now active rotations of the particle, which we call active spinning motions, are taken into account. They are induced by setting $\zeta$ to a positive value, $\zeta=1.5$, in Eq.~(\ref{eq:Omega}). 
Since the rotational symmetry of space is broken by the shear flow given by Eq.~(\ref{eq:ugamma}) with $\dot{\gamma}>0$, we distinguish between two cases. First, we consider clockwise active rotations of the particle, after that counter-clockwise spinning motions. The rotational part of the shear flow itself is oriented in the clockwise direction. 
Our results are again presented in terms of the quantities introduced in Eqs.~(\ref{S}), (\ref{v}), and (\ref{psi}).

\subsection{Clockwise active rotation}

\begin{figure*}[tbp]
  \begin{center}
         \includegraphics[width=7 in]{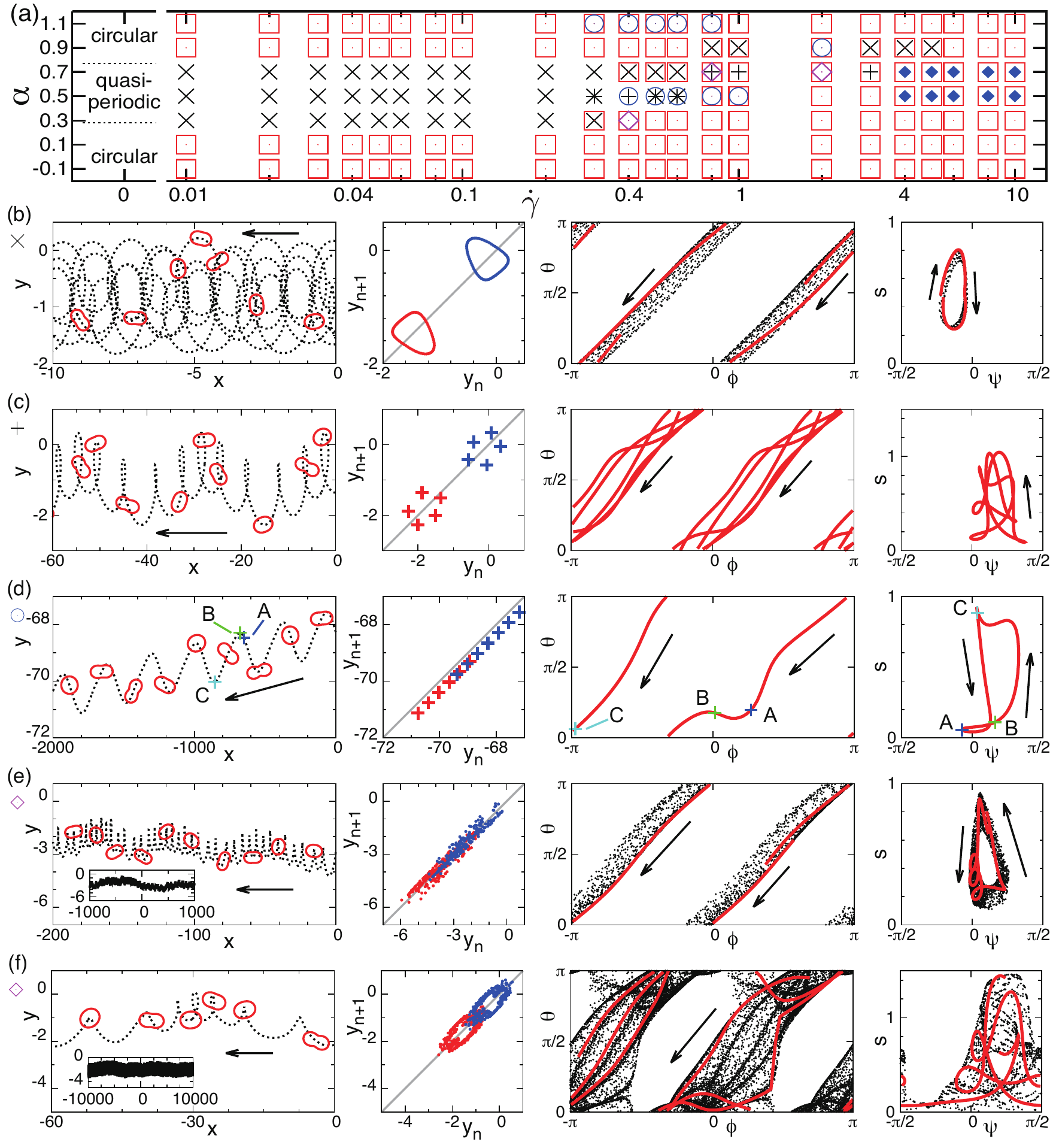}
      \caption{
      (a) Dynamical phase diagram and (b)--(f) trajectories in real space ($1$st column), return maps ($2$nd column), attractors in $\theta$-$\phi$ 
      space ($3$rd column) as well as in $s$-$\psi$ space ($4$th column) of the 
      typical dynamical motions, obtained by solving Eqs.~(\ref{eq:x}) and (\ref{eq:v})--(\ref{eq:Omega}) numerically in two dimensions for clockwise active rotations ($\zeta=1.5$); 
      (b)       {\it quasi-periodic motion} 
      for $\alpha=0.5$ and $\dot{\gamma}=0.1$ indicated by the black crosses in panel~(a); 
      (c)       {\it periodic motion} 
      for $\alpha=0.7$ and $\dot{\gamma}=1$ marked by the black plus symbols in panel~(a); 
      (d)       {\it winding II motion} 
      for $\alpha=0.5$ and $\dot{\gamma}=0.8$ identified by the blue open circles in panel~(a); 
      (e) and (f)       {\it chaotic motions} 
      for $\alpha=0.3$ and $\dot{\gamma}=0.4$ as well as for $\alpha=0.7$ and $\dot{\gamma}=2$, respectively, marked by the purple open diamonds in panel~(a). 
      Arrows in panels~(b)--(f) show the directions of motion. 
      Some snapshots of the particle, the size of which is adjusted for illustration, are superimposed to the trajectory in real space. 
      In panels~(e) and (f), insets show the corresponding trajectories over longer time intervals. 
      The trajectories in $\theta$-$\phi$ and $s$-$\psi$ phase space in panels (b), (e), and (f) are indicated by the black points. We also include short-time trajectories as red lines. 
      Red open squares stands for the       {\it cycloidal I motion} 
      as already characterized in Fig.~\ref{fig:zeta-0.1}(c) for $\zeta=-0.1$. 
      The blue filled diamonds in panel~(a) represent an       {\it undulated cycloidal I motion} 
      as it is further illustrated by Fig.~\ref{fig:zeta1.5_positive}(c). 
      Superimposed symbols in panel~(a) indicate the observation of different trajectory types depending on the initial conditions. 
      } \label{fig:zeta1.5_negative}
  \end{center}
\end{figure*}

Without an externally imposed shear flow, i.e.\ for $\dot{\gamma}=0$, the situation of active spinning has been recently investigated by some of the present authors 
\cite{Tarama2012,Tarama2013}. 
For the parameters that we have chosen in this paper, two types of motion have been found in the absence of the shear flow: circular and quasi-periodic motions. We repeat these results on the left border of our phase diagram Fig.~\ref{fig:zeta1.5_negative}(a) in the column $\dot{\gamma}=0$. There, with increasing self-propulsion strength $\alpha$, the circular motion is reentrant 
\cite{Tarama2012,Tarama2013}. 

With the shear flow now turned on and an active spinning in clockwise direction, the circular motion changes to the corresponding cycloidal I motion that was already obtained in the previous section and characterized in Fig.~\ref{fig:zeta-0.1}(c) for $\zeta=-0.1$. It covers a major part of our phase diagram Fig.~\ref{fig:zeta1.5_negative}(a) and is indicated by the red open squares. 
A cycloidal trajectory naturally results, when advection due to the flow is superimposed to a circular motion. 

Next, a {\it quasi-periodic motion} 
is marked by the black crosses in the phase diagram Fig.~\ref{fig:zeta1.5_negative}(a). It occurs at intermediate self-propulsion strengths $\alpha$. Interestingly, this type of motion is suppressed with increasing shear-rate $\dot{\gamma}$. We further characterize it in Fig.~\ref{fig:zeta1.5_negative}(b) for $\alpha=0.5$ and $\dot{\gamma}=0.1$. Obviously, the motion is not simply periodic as evident from the real-space trajectory and from the trajectories indicated in the $\theta$-$\phi$ and $s$-$\psi$ phase spaces by the black dots. However, it is quasi-periodic and not chaotic, because the return maps give discrete closed loops as shown in the second column of Fig.~\ref{fig:zeta1.5_negative}(b). 
The difference between the quasi-periodic motion in the absence of the shear flow and the one in the presence of the shear flow is simply that the particle stays within a finite area in the former case while in the latter case it escapes over time in the positive or negative $x$-direction.

When the shear rate $\dot{\gamma}$ is increased, several new types of motion are found that we have not observed before. Interestingly, all of them are sensitive to the initial conditions. We find a coexistence of at least two types of motion at every point of the phase diagram that we investigated for these new dynamic states, which leads to the superposition of the symbols in Fig.~\ref{fig:zeta1.5_negative}(a).  

First, at higher shear rates, a {\it periodic motion} 
that cannot be observed at low shear rates is found at some positions in the phase diagram. This dynamic state is marked by the black plus symbols in Fig.~\ref{fig:zeta1.5_negative}(a) and further illustrated in Fig.~\ref{fig:zeta1.5_negative}(c) for $\alpha=0.7$ and $\dot{\gamma}=1$. 
The real-space trajectory 
appears as a commensurately modulated cycloid. We can distinguish this kind of motion from the quasi-periodic motion by the return map in the second column of Fig.~\ref{fig:zeta1.5_negative}(c), where thousands of measured trajectory extrema condense on ten discrete points in contrast to the closed loop object in Fig.~\ref{fig:zeta1.5_negative}(b). There are even some coexistence points of periodic and quasi-periodic motion in the phase diagram, induced by different initial conditions. 

Next, in analogy to the winding I motion of the previous section, a {\it winding II motion} 
is identified at the positions of the blue open circles in the phase diagram Fig.~\ref{fig:zeta1.5_negative}(a). We characterize it in Fig.~\ref{fig:zeta1.5_negative}(d) for $\alpha=0.5$ and $\dot{\gamma}=0.8$. 
Since the particle in real space continuously descends in $y$-direction, the discrete points in the return map descend along the diagonal. 
The winding II motion can easily be distinguished from its counterpart, the winding I motion in Fig.~\ref{fig:zeta-0.1}(e), by the trajectory in $\theta$-$\phi$ phase space. When observed from the laboratory frame, the particle features full rotations of its long axis of deformation 
$\mathbf{S}$ in the winding II state, while only oscillations of this long axis occur 
in the winding I state. 
To facilitate the connection between the real- and phase-space trajectories in Fig.~\ref{fig:zeta1.5_negative}(d), we marked corresponding points by the capital letters ``A'', ``B'', and ``C''. 
We found a three-state coexistence region including the winding II motion, the periodic motion, and the quasi-periodic motion in the phase diagram Fig.~\ref{fig:zeta1.5_negative}(a) around the point $\alpha=0.5$ and $\dot{\gamma}=0.6$.

Most interestingly, we now also find {\it chaotic states} 
of the dynamic behavior of our single deformable active particle subjected to linear shear flow. In the phase diagram Fig.~\ref{fig:zeta1.5_negative}(a) they are marked by the purple open diamonds. 
Figures \ref{fig:zeta1.5_negative}(e) and (f) 
show the characteristics of two chaotic dynamic states for $\alpha=0.3$ and $\dot{\gamma}=0.4$ as well as for $\alpha=0.7$ and $\dot{\gamma}=2$, respectively. 
The inset figures in the real-space trajectory plots in the first column of Figs.~\ref{fig:zeta1.5_negative}(e) and (f) depict the trajectory over longer time intervals. 
Closer inspection shows that the attractors in the $\theta$-$\phi$ and $s$-$\psi$ phase spaces of Fig.~\ref{fig:zeta1.5_negative}(e) are similar to the ones of the quasi-periodic motion in Fig.~\ref{fig:zeta1.5_negative}(b). However, the return maps are different enough to distinguish these two separate types of motion: 
the return map of the quasi-periodic motion forms a simple closed loop, while that of the chaotic motion in Fig.~\ref{fig:zeta1.5_negative}(e) is dispersed around the diagonal $y_{n+1}=y_n$. 
Likewise, we can distinguish the chaotic motion in Fig.~\ref{fig:zeta1.5_negative}(f) from the undulated cycloidal I motion discussed below in Fig.~\ref{fig:zeta1.5_positive}(c) via their return maps, although the attractors in $\theta$-$\phi$ space and $s$-$\psi$ space are similar.

At large shear rates $\dot{\gamma}$ and intermediate self-propulsion strengths $\alpha$, another dynamic state was observed. It is marked by the blue filled diamonds in the phase diagram Fig.~\ref{fig:zeta1.5_negative}(a) and we call it an {\it undulated cycloidal I motion}. 
Since it also appears in the case of counter-clockwise active rotations of the particle, we discuss it below together with the dynamic states observed in that case.

\subsection{Counter-clockwise active rotation}

\begin{figure*}[tbhp]
  \begin{center}
         \includegraphics[width=7 in]{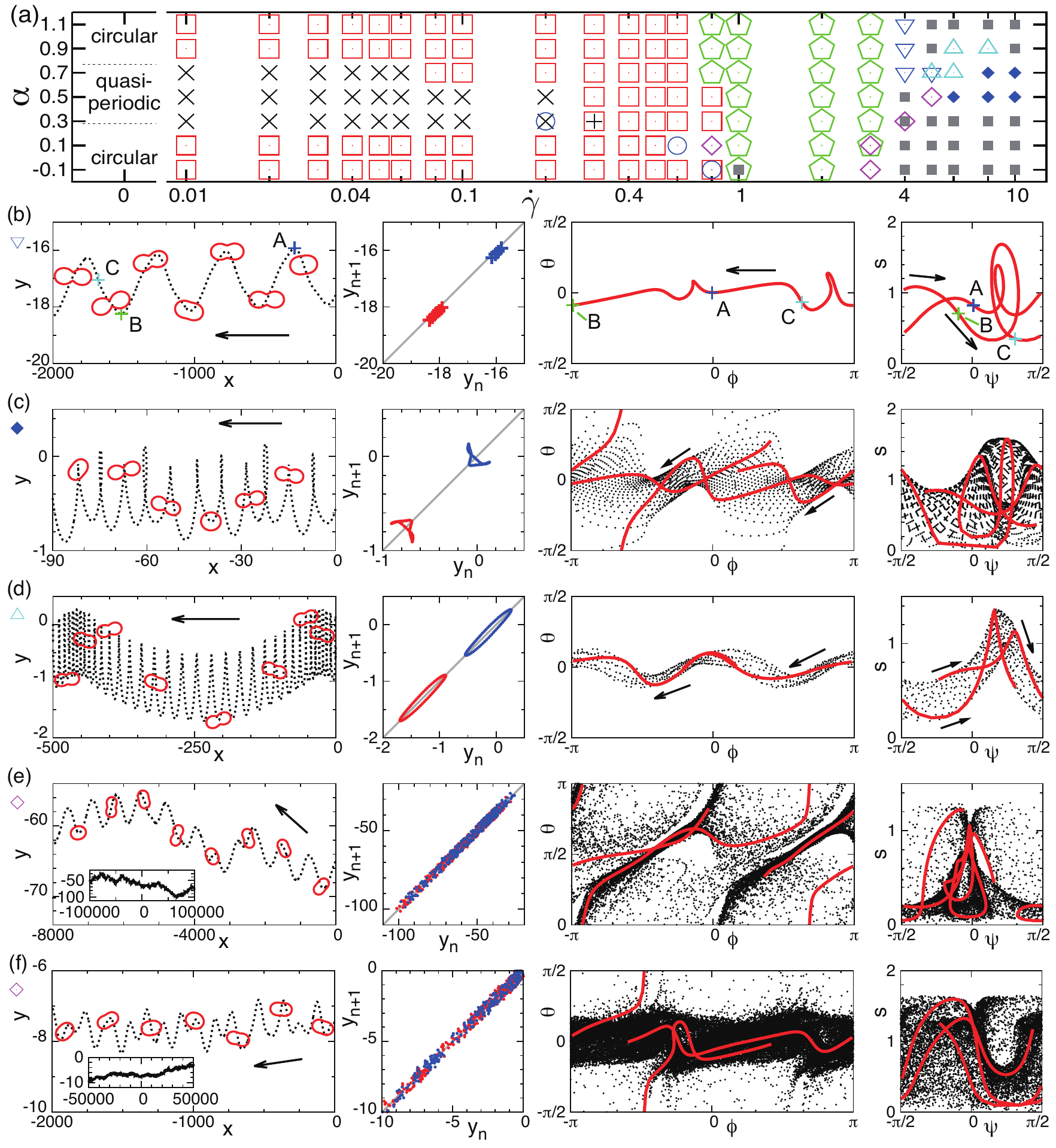}
      \caption{
      (a) Dynamical phase diagram and (b)--(f) trajectories in real space ($1$st column), return maps ($2$nd column), attractors in $\theta$-$\phi$ 
      space ($3$rd column) as well as in $s$-$\psi$ space ($4$th column) of the 
      typical dynamical motions, obtained by solving Eqs.~(\ref{eq:x}) and (\ref{eq:v})--(\ref{eq:Omega}) numerically in two dimensions for counter-clockwise active rotations ($\zeta=1.5$);  
      (b)       {\it winding III motion} 
      for $\alpha=0.9$ and $\dot{\gamma}=4$ indicated by the blue open downward triangles in panel~(a); 
      (c)       {\it undulated cycloidal I motion} 
      for $\alpha=0.5$ and $\dot{\gamma}=6$ specified by the blue filled diamonds in panel~(a); 
      (d)       {\it undulated cycloidal II motion} 
      for $\alpha=0.9$ and $\dot{\gamma}=6$ marked by the blue open upward triangles in panel~(a); 
      (e) and (f)       {\it chaotic motions} 
      for $\alpha=0.1$ and $\dot{\gamma}=0.8$ as well as for $\alpha=0.3$ and $\dot{\gamma}=4$, respectively, indicated by the purple open diamonds in panel~(a). 
      Arrows in panels~(b)--(f) show the directions of motion. 
      Some snapshots of the particle, the size of which is adjusted for illustration, are superimposed to the trajectory in real space. 
      In panels~(e) and (f), insets show the corresponding trajectories over longer time intervals. 
      The trajectories in $\theta$-$\phi$ and $s$-$\psi$ phase space in panels (c)--(f) are indicated by the black points. We also include short-time trajectories as red lines. 
      The dynamic states corresponding to the other symbols in panel~(a) that are not further characterized in panels (b)--(f) have already been explained in Figs.~\ref{fig:zeta-0.1} and \ref{fig:zeta1.5_negative}. 
      Superimposed symbols in panel~(a) indicate the observation of different trajectory types depending on the initial conditions. 
      } \label{fig:zeta1.5_positive}
  \end{center}
\end{figure*}

Finally, we analyze the case of counter-clockwise spinning motions of the active particle, i.e.\ active rotations in the direction opposite to that of the rotational part of the shear flow. Without the shear flow at $\dot{\gamma}=0$, the rotational symmetry in space is not broken, and the dynamical states of clock- and counter-clockwise rotations are identical (except for the sense of rotation). 

At low shear rates, the dynamics for both senses of rotation is still similar as can be inferred when comparing the corresponding phase diagrams 
Figs.~\ref{fig:zeta1.5_negative}(a)
 and 
\ref{fig:zeta1.5_positive}(a)
for low values of $\dot{\gamma}$. Again, a cycloidal I motion appears for both high and low self-propulsion strengths $\alpha$. Likewise, a quasi-periodic 
motion emerges at intermediate self-propulsion strengths $\alpha$. They are marked by the red open squares and black crosses 
in the phase diagram Fig.~\ref{fig:zeta1.5_positive} and were discussed in Figs.~\ref{fig:zeta-0.1}(d) and \ref{fig:zeta1.5_negative}(b), respectively. 
Increasing the shear rate $\dot{\gamma}$, the quasi-periodic motion becomes unstable in favor of the cycloidal I motion. Also the periodic motion, indicated by the black pluses 
and previously characterized in Fig.~\ref{fig:zeta1.5_negative}(c), as well as the winding II motion, marked by the blue open circle and previously depicted in Fig.~\ref{fig:zeta1.5_negative}(d), are recovered. Coexistence of different dynamic states again occurs and is shown by the superposition of different symbols in the phase diagram Fig.~\ref{fig:zeta1.5_positive}(a). 

Interestingly, at 
large
shear rates $\dot{\gamma}\gtrsim 1$, we observe an active-straight motion at all investigated self-propulsion strengths $\alpha$. It is indicated in the phase diagram Fig.~\ref{fig:zeta1.5_positive}(a) by the green open pentagons and was previously discussed in Fig.~\ref{fig:zeta-0.1}(b). 
The origin of the emergence of the active straight motion at these shear rates can be easily understood: it appears when the rotation due to the active spinning motion in the counter-clockwise direction and the rotation due to the external flow in the clockwise direction balance each other.
 
At still larger shear rates $\dot{\gamma}$, this balance is no longer maintained and different types of motion appear. At high self-propulsion strength $\alpha$, the particle next undergoes a {\it winding III motion} 
as denoted by the blue open downward triangles in the phase diagram Fig.~\ref{fig:zeta1.5_positive}(a). This motion is characterized in Fig.~\ref{fig:zeta1.5_positive}(b) for $\alpha=0.9$ and $\dot{\gamma}=4$. 
In contrast to the winding I and winding II motions in Figs.~\ref{fig:zeta-0.1}(e) and \ref{fig:zeta1.5_negative}(d), respectively, the angle $\theta$ as viewed from the laboratory frame always remains of small magnitude with values close to zero. This means that the particle always remains elongated along the horizontal direction. Only small oscillations of the long axis of deformation  
occur that are due to the competition between the active rotation in the counter-clockwise direction and the clockwise rotation induced by the shear flow. 
In Fig.~\ref{fig:zeta1.5_positive}(b), capital letters ``A'', ``B'', and ``C'' are again used to mark corresponding points along the trajectories in real space, in $\theta$-$\phi$ phase space, and in $s$-$\psi$ phase space. 

In Sec.~\ref{sec:zeta=-0.1} we have already found a dynamic mode that features an almost horizontal elongation of the particle at all times. It was the cycloidal II motion, obtained without active spinning, and depicted in Fig.~\ref{fig:zeta-0.1}(f). Indeed, we find this type of motion again when increasing the shear rate $\dot{\gamma}$ from the winding III motion at high self-propulsion strengths $\alpha$. In addition to that, it is also the dominant dynamic mode at large shear rate $\dot{\gamma}$ but small self-propulsion strength $\alpha$. We indicate it again by the gray filled squares in Fig.~\ref{fig:zeta1.5_positive}(a). 

Apart from the cycloidal I and II motions, depicted previously in Figs.~\ref{fig:zeta-0.1}(c) and (d) as well as in Fig.~\ref{fig:zeta-0.1}(f), respectively, there are two other types of cycloidal motions. 
One of them is the undulated cycloidal I motion that has already been found for clockwise active spinning motion. It is marked by the blue filled diamonds in the phase diagrams Figs.~\ref{fig:zeta1.5_negative}(a) and \ref{fig:zeta1.5_positive}(a). We now further characterize it in Fig.~\ref{fig:zeta1.5_positive}(c) for $\alpha=0.5$ and $\dot{\gamma}=6$. 
An undulation of the cycloidal amplitude is apparent from the real-space trajectory as well as from the return map. As can be seen in $\theta$-$\phi$ phase space, the long axis of deformation 
makes full rotations in the laboratory frame. 

The other further cycloidal type is the new {\it undulated cycloidal II motion} 
that we find only for counter-clockwise particle spinning and that we mark 
by the blue open upward triangles in Fig.~\ref{fig:zeta1.5_positive}(a). We illustrate this dynamic mode in Fig.~\ref{fig:zeta1.5_positive}(d) for $\alpha=0.9$ and $\dot{\gamma}=6$. In contrast to the undulated cycloidal I motion, there is no full rotation of the long axis of deformation 
in the laboratory frame as becomes obvious in $\theta$-$\phi$ phase space.

Again we also observe chaotic motions, which are represented by the purple open diamonds in Fig.~\ref{fig:zeta1.5_positive}(a). 
Characteristics of these chaotic motions are displayed in Figs.~\ref{fig:zeta1.5_positive}(e) and (f) for $\alpha=0.1$ and $\dot{\gamma}=0.8$ as well as for $\alpha=0.3$ and $\dot{\gamma}=4$, respectively. 
A qualitative difference between the two depicted chaotic motions becomes obvious from the plots in phase space. While in the first case of Fig.~\ref{fig:zeta1.5_positive}(e) the long axis of deformation 
of the particle tends to rotate together with the velocity direction, it has a tendency to remain horizontal in the second case of Fig.~\ref{fig:zeta1.5_positive}(f). Both tendencies can be inferred from the 
dark bands in the $\theta$-$\phi$ plots. 
Generally, we find that the trajectories in phase space in Figs.~\ref{fig:zeta1.5_positive}(e) and (f) are more delocalized than for the chaotic states of clockwise rotations that we have illustrated in Figs.~\ref{fig:zeta1.5_negative}(e) and (f).

\section{Summary and Conclusions} \label{sec:discussion}

In this paper, we have investigated the dynamics of a deformable active particle in shear flow. 
For that purpose, we have considered a soft deformable particle with two types of activity: 
one is a spontaneous propulsion and the other one is a spontaneous spinning motion. 
The deformation of the particle is described by a symmetric traceless tensor variable, and its rotation by an anti-symmetric tensor variable. 
Further variables are the position of the center of mass and its relative velocity with respect to the flow. The externally imposed linear shear flow is included by taking into account its deformational and its rotational impact. Using symmetry arguments, we derive coupled dynamic equations for all of these variables. 
Our equations reduce to  known models in the two limits of vanishing shear flow
and vanishing particle deformability. On the one hand, in the limit of vanishing shear flow, we reproduce the previous results of Refs.~\citenum{Tarama2012,Tarama2013}. 
On the other hand, for vanishing particle deformability, we obtain an approximate analytical solution that is consistent with previous investigations 
\cite{Hagen2011}.

Various 
types of motion arise  
as numerical solutions of the full set of dynamical equations including active straight motion, periodic motions, motions on regular and undulated cycloids, winding motions, as well as quasi-periodic and chaotic motions
induced at high shear rates. 
In order to characterize and distinguish these dynamical states, we have analyzed and categorized them via their trajectories, corresponding return maps, as well as their attractors in phase space. Also the two situations of clockwise and counter-clockwise rotations 
with respect to the direction of the shear flow are distinguished and lead to partially different results, in particular at high shear rates.

Our predictions can be verified in experiments 
on self-propelled droplets exposed to
shear flow.
For instance, in some experiments \cite{Nagai2005,Takabatake2011} self-propelled droplets on liquid-air interfaces 
can be exposed to linear shear fields by putting the 
carrier liquid 
between two 
parallel confining
walls that move 
alongside
into opposite directions. 
This induces an approximately planar linear shear gradient at the surface of the carrier liquid, if the liquid container is sufficiently deep.
Since the 
motion of the droplets is confined to the liquid-air interface, 
 the geometry is quasi-two dimensional. 
Using this experimental set-up, it is in principle possible to verify the phenomena predicted by our analysis.

Future studies should address several extensions of our model: first of all, different prescribed flow 
fields can be explored using our equations.  Most noticeable examples include
 a Poiseuille flow \cite{Stark2012} or an imposed
vorticity field. We expect again a manifold of different types of motion in these flow fields presuming that
the different flow topologies will induce different types of motion. The next step is to extend our 
analysis to a finite concentration of particles and to include steric interactions between them.
This is a complex problem which is already very difficult for rigid self-propelled particles
\cite{Vicsek1995,Toner1995,Simha2002,Sokolov2007,Ginelli2010,Itino2011,Menzel2012,Wensink2008,Wensink2012}.
Another step is to extend the current analysis of the model to three spatial dimensions. In a previous study, some of us demonstrated that -- without shear flow -- a particle can exhibit additional qualitatively different types of motion when comparing a three- to a two-dimensional set-up \cite{Tarama2013}. In the case without shear flow, these were additional helical and superhelical types of motion \cite{Tarama2013}. Thus we expect that further new types of dynamics can arise in three dimensions when the shear flow is included.
Finally one could access the dynamics of propelled vesicles by using our analysis as a starting point.
Here one should impose the constraints of constant volume and constant surface area of the deformable particle.
The dynamics of passive vesicles in shear flow has been explored quite extensively in recent years
\cite{Lipowsky,Noguchi} with various specific effects like
tank-treading motion, lifting \cite{Seifert,Cantat}, wrinkling \cite{Finken}, tumbling and swinging
 \cite{Gompper,Finken2}. It would indeed be interesting to generalize all these effects to
 self-propelled vesicles.

\acknowledgements

This work was supported by the JSPS Core-to-Core Program ``International research network for non-equilibrium dynamics of soft matter'' 
and ``Non-equilibrium dynamics of soft matter and information,'' 
and by 
 a Grant-in-Aid for Scientific Research C (No. 23540449) from JSPS and a Grant-in-Aid for Scientific Research A (No.~24244063) from Mext. 
M.T.\ is supported by the Japan Society for the Promotion of Science Research Fellowship for Young Scientists. A.M.M.\ and H.L.\ gratefully acknowledge support from the Deut\-sche For\-schungs\-ge\-mein\-schaft through the German--Japanese project ``Nicht\-gleich\-ge\-wichts\-ph\"a\-no\-me\-ne in Wei\-cher Ma\-te\-rie/Soft Matter'' No.~LO 418/15. R.W.\ gratefully acknowledges financial support from a Postdoctoral Research Fellowship (Grant No. WI 4170/1-1) of 
the Deut\-sche For\-schungs\-ge\-mein\-schaft.


\end{document}